" **Impedance-Based Stability Analysis for Interconnected Converter Systems with Open-Loop RHP Poles** "

Yicheng Liao and Xiongfei Wang

This is a preprint of the paper submitted to the IEEE Transaction on Power Electronics.

# Impedance-Based Stability Analysis for Interconnected Converter Systems with Open-Loop RHP Poles

Yicheng Liao, *Student Member, IEEE*, Xiongfei Wang, *Senior Member, IEEE*

(Corresponding author: Xiongfei Wang)

*Abstract* – **Small-signal instability issues of interconnected converter systems can be addressed by the impedance-based stability analysis method, where the impedance ratio at the point of common connection of different subsystems can be regarded as the open-loop gain, and thus the stability of the system can be predicted by the Nyquist stability criterion. However, the right-half plan (RHP) poles may be present in the impedance ratio, which then prevents the direct use of Nyquist curves for defining stability margins or forbidden regions. To tackle this challenge, this paper proposes a general rule of impedance-based stability analysis with the aid of Bode plots. The method serves as a sufficient and necessary stability condition, and it can be readily used to formulate the impedance specifications graphically for various interconnected converter systems. Experimental case studies validate the correctness of the proposed method.**

I. INTRODUCTION

In recent years, the increasing use of power electronic converters has brought stability issues to the modern power systems, such as the sideband oscillations and harmonic oscillations [1]. The impedance-based stability method provides an attractive approach for addressing the stability issue of interconnected converter systems. The basic idea is that the impedance ratio of the two subsystems, divided at a given point of common connection (PCC), can serve as the open-loop gain of the whole system, through which the system dynamics can be assessed based on the Nyquist stability criterion (NSC). This method only concerns the frequency responses of the terminal impedances of different subsystems, and thus is flexible in different scenarios, especially for "black box" systems whose internal parameters are unknown.

A comprehensive application of this method dates from 1976 by Undrill and Kostyniak, where the sub-synchronous oscillations are analyzed based on the impedances of the generator and the transmission network [2]. In the same year, Middlebrook first applied the method to design the input filters of dc-dc converters [3], where the system stability is guaranteed by confining the minor loop gain (defined as the ratio of the filter impedance and the converter input impedance) within the unity circle on the complex plane. Later on, the minor loop gain was defined as the source-load impedance ratio for the stability analysis of distributed power systems (DPSs). By preventing the minor loop gain from entering the so-called forbidden regions, Wildrick et al. developed a method for impedance specifications with the aid of Bode plots to achieve a stable DPS, also known as the gain margin & phase margin (GMPM) criterion [4]. This approach provides an intuitive way to shape the impedances of converters for meeting the pre-defined stability margin.

Based on the concept of forbidden region, several stability criteria for DPSs were proposed, such as the opposing argument criterion (OPAC) for individual load impedance specifications [5]-[6], the ESAC criterion [7], the root exponential stability criterion (RESC) [8], and the maximum peak criterion for robust stability assessment [9]. Yet, those criteria are all sufficient conditions. Li and Zhang developed a necessary and sufficient stability criterion (NSSC) by reducing the forbidden region as the real axis at the left of $(-1, j0)$ [10]. Based on the NSSC, the impedance profiles can also be specified on Bode plots to ensure stability. However, all the stability criteria that are based on the source-load impedance ratio may confront obstacles in some cases. For current source connected systems (CSCSs), as the current source operates with a stable and low output admittance, the load-source impedance ratio is utilized instead [11]-[12], such that the right-half plane (RHP) poles will never exist in the impedance ratio. Alternatively, the inverse NSC can be applied to the original source-load impedance ratio, yet the possible RHP poles present in the current source impedance should be carefully dealt with, since they will become the open-loop RHP poles for the source-load impedance ratio [13]. If open-loop RHP poles exist, the Nyquist plot has to encircle the unity circle for ensuring the system stability. Consequently, the forbidden-region-based impedance shaping and design cannot work.

There have been a few methods proposed to avoid the impacts of open-loop RHP poles [14]-[17]. Liu et al. categorized all interconnected systems into Z+Y systems and Z+Z systems, and then developed the corresponding stability criteria for different systems [14]-[15]. For Z+Y systems [14], where one subsystem has a stable terminal impedance (Type Z), and the other has a stable terminal admittance (Type Y), the RHP poles cannot be present in the impedance ratio of Type Z system over Type Y system. Therefore, such a ZY ratio can be regarded as the minor loop gain for stability analysis, and the previous forbidden regions can be applied. Both DPSs and CSCSs are Z+Y systems, where the impedance ratio-based method works well. Zhang et al. applied this method in a DPS with multiple converters, where an aggregation method was introduced by lumping all the Type Z converters, which are called as bus voltage controlled converters together, and aggregating all the Type Y converters, which are named as bus current controlled converters together. Thus, the impedance ratio of Z+Y systems can be used for the stability analysis [16]. Nevertheless, in some cases, e.g. hybrid energy storage systems (HESSs), each subsystem only has a stable impedance, which are called as Z+Z systems. The RHP poles may be produced in the impedance ratio if there are RHP zeros in the denominator impedance. To avoid the presence of RHP poles, an impedance sum criterion was developed by checking the number of RHP zeros in the impedance sum, which is based on the argument principle, since it serves as the characteristic equation of the whole system [15]. However, the interactions between different impedances are intangible in this method, since the impedance sum has to be calculated. Moreover, it was reported in [17] that the multi-loop NSC provides an alternative impedance-based stability analysis method for systems with open-loop RHP poles. Yet, it is not easy to open the multiple loops in practical systems if the analytical model of the system is unknown.

Although the impedance-ratio-based method is advantageous for analyzing the stability of interconnected converter systems, there exist some gaps:

1) There is no general approach to formulate the impedance ratio for the stability analysis. The current practice is to define the impedance ratio based on the source types of converters, or by avoiding the RHP poles, but it is not applicable for Z+Z systems.

2) The forbidden region-based methods that excel in the design-oriented analysis with specific impedance profiles, fail in the stability analysis if the impedance ratio has RHP poles.

3) Although the NSC can be used to analyze the stability of all kinds of systems, the impedance ratio has

to be considered as a whole on the Nyquist plot. It is hard to investigate the interactions between different impedances through the Nyquist plot and to design the system by impedance specifications.

To fill in these gaps, this paper proposes a general rule for the impedance-based stability analysis on Bode plots, which is equivalent to the NSC, yet enables to formulate impedance specifications even if open-loop RHP poles exist. The rest of the paper is organized as follows: Section II reviews the present impedance-based stability analysis methods; Section III proposes the general rule for impedance-based stability analysis, and its advantages are given through comparison; Section IV provides the experimental case studies for verification; Section V finally draws the conclusions.

## II. Review of Impedance-Based Stability Analysis

The conventional impedance-based stability analysis methods are critically reviewed in this section. For an interconnected system, it can be regarded as two subsystems connected at the PCC. Each subsystem can be represented by a voltage source in series with an impedance or a current source in parallel with an impedance, according to the Thevenin's theorem or the Norton's theorem, as shown in Fig. 1. Then the voltages at the PCCs in Fig. 1(a) and (b) can be, respectively, expressed as

$$V = \frac{Z_2}{Z_1+Z_2}V_1 + \frac{Z_1}{Z_1+Z_2}V_2 = \frac{1}{1+\frac{Z_1}{Z_2}}V_1 + \frac{\frac{Z_1}{Z_2}}{1+\frac{Z_1}{Z_2}}V_2 = \frac{1}{1+\frac{Z_2}{Z_1}}V_1 + \frac{\frac{Z_2}{Z_1}}{1+\frac{Z_2}{Z_1}}V_2, \quad (1)$$

$$V = \frac{Z_1 Z_2}{Z_1+Z_2}I_1 + \frac{Z_1 Z_2}{Z_1+Z_2}I_2 = \frac{Z_1}{1+\frac{Z_1}{Z_2}}I_1 + \frac{Z_1}{1+\frac{Z_1}{Z_2}}I_2 = \frac{Z_2}{1+\frac{Z_2}{Z_1}}I_1 + \frac{Z_2}{1+\frac{Z_2}{Z_1}}I_2. \quad (2)$$

Both Eqs. (1) and (2) are established in the *s* domain with "*s*" omitted for brevity. It can be seen that $Z_1+Z_2$ can be regarded as the characteristic equation of the whole system, and both the impedance ratios $Z_1/Z_2$ and $Z_2/Z_1$ can be regarded as the open-loop gain. Therefore, the stability of the interconnected system relies significantly on the terminal impedances of the two subsystems. The conventional impedance-based stability analysis methods can be mainly categorized into three types, the impedance-ratio-based method, the impedance-sum-based method, and the NSC for multi-loop systems.

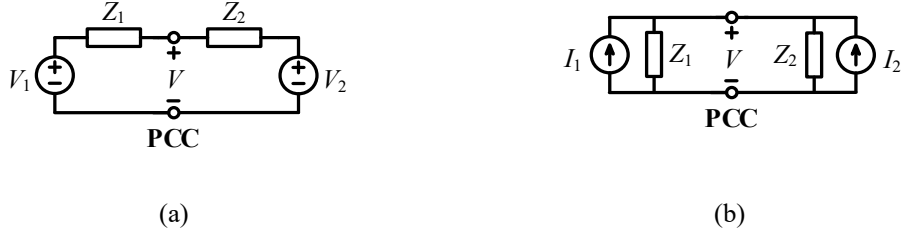

Fig. 1 An interconnected system. (a) Thevenin's equivalence; (b) Norton's equivalence.

*A. Impedance-ratio-based method*

For the impedance-ratio-based method, the impedance ratio $Z_1/Z_2$ or $Z_2/Z_1$ can be regarded as the open-loop transfer function (or the minor-loop gain) of the interconnected system. Nyquist proposed the regeneration theory, also known as the NSC nowadays, to assess the system stability only based on the open-loop transfer function of a system [18]. According to the NSC [19], the system is stable if and only if the Nyquist plot of the impedance ratio encircles the critical point $(-1, j0)$ on the complex plane $N$ times anti-clockwise, where $N$ equals to the number of RHP poles in the impedance ratio.

*a) Stability analysis based on the NSC*

Since the impedance ratio is defined manually, the open-loop RHP poles may exist, and the open-loop transfer function may be an improper function, where the degree of the numerator is greater than the degree of the denominator. Consequently, the utilization of the NSC is dependent on the formulation of the impedance ratio. Table I summarizes the applications of the NSC or the inverse NSC [20] based on the formulation of the impedance ratio, where the impedance ratio $Z_1/Z_2$ is assumed for analysis. It can be seen from Table I that the inverse NSC is essentially another form of the NSC.

Table I  Applications of the NSC based on the definition of impedance ratio

| Methods | NSC [19] | Inverse NSC [20] |
|---|---|---|
| Stable condition | $\mathcal{N}[Z_1/Z_2] = -\mathcal{P}[Z_1/Z_2]$ | $\mathcal{N}[1/(Z_1/Z_2)] = -\mathcal{Z}[Z_1/Z_2]$ |
| Features | Impedance ratio defined as the "minor loop gain" ($|Z_1| < |Z_2|$) | Impedance ratio defined as the inverse "minor loop gain" ($|Z_1| > |Z_2|$) |
| Cases | Input filter design of converters [3], source-load impedance ratio for DPSs [4]-[8], load-source impedance ratio for CSCSs [11]-[12] | Source-load impedance ratio for CSCSs [13] |
| Constraints | 1) with prior knowledge of the number of open-loop RHP poles  2) failure in impedance interaction investigation and specification  3) lack of frequency information | |

Notes: $\mathcal{N}$ denotes the number of Nyquist encirclements around $(-1, j0)$, $\mathcal{P}$ denotes the number of RHP poles, and $\mathcal{Z}$ denotes the number of RHP zeros.

The NSC or the inverse NSC can be checked on the Nyquist plot of the impedance ratio, which provides a sufficient and necessary condition for the stability assessment. However, there are some constraints on the direct use of these methods:

1) A prior knowledge of the number of RHP poles of the impedance ratio is required, and hence additional identifications of RHP zeros in the denominator impedance are needed.

2) The impedance ratio has to be treated as a whole on the Nyquist plot, thus the interactions between different impedances cannot be further studied.

3) The frequency information is not visible in the Nyquist plot. Even though the stability can be analyzed accurately, it provides little insight into the design of converter controllers.

*b) Stability analysis based on forbidden regions*

According to the constraints of the NSC-based methods, some efforts have been done for facilitating the stability analysis, which are based on forbidden regions. It is preferable to formulate the impedance ratio without RHP poles, thus the system is stable if and only if the Nyquist plot of $Z_1/Z_2$ does not encircle $(-1,$ $j0$). This precondition can be easily realized for the Z+Y systems [14], where one subsystem has a stable terminal impedance, and the other one has a stable terminal admittance. Based on this, the open-loop gain (also named as the return ratio transfer function) is defined as

$$L = \frac{Z_{\text{Type\_Z}}}{Z_{\text{Type\_Y}}} = Z_{\text{Type\_Z}} \cdot Y_{\text{Type\_Y}}, \tag{3}$$

in which, both $Z_{\text{Type\_Z}}$ and $Y_{\text{Type\_Y}}$ have no RHP poles, thus the whole interconnected system has no open-loop RHP poles. For DPSs, the voltage sources can be regarded as Type Z systems, and the loads can be regarded as Type Y systems. For CSCSs, the current sources can be regarded as Type Y systems, and the grid or other loads can be regarded as Type Z systems. For a more complicated system with multiple converters, the different subsystems can be aggregated based on the system types. It was reported in [16] that for a DPS, all the bus voltage controlled converters (BVCCs) should be lumped together as the numerator impedance, and all the bus current controlled converters (BCCCs) should be aggregated together as the denominator impedance, since the BVCCs are Type Z systems, and the BCCCs are Type Y systems.

Based on the precondition of zero open-loop RHP poles, several forbidden regions were proposed to specify different stability margins [3]-[8], [10], [21], as shown in the shaded areas in Fig. 2, where GM and PM denote the gain margin (dB) and phase margin (degree), respectively. gm is the gain margin in the real coordinate, which is

$$\text{gm} = 10^{\frac{\text{GM}}{20}}. \tag{4}$$

If the impedance ratio is confined out of the forbidden regions, the system stability will be guaranteed. However, it is noted that the impedance ratio that enters into the forbidden regions does not imply an unstable system. Those criteria based on forbidden regions are sufficient stability conditions. The smaller the forbidden region is, the less conservative the criterion is.

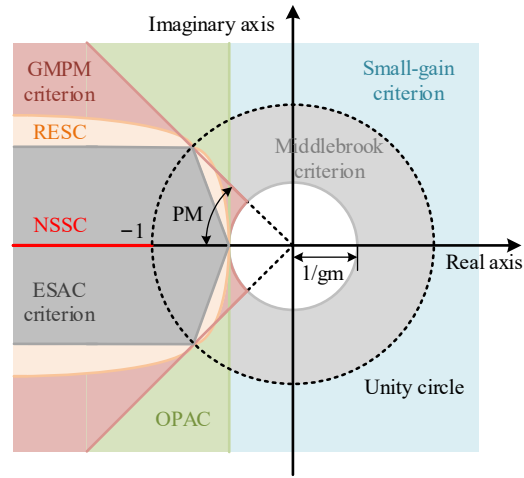

Fig. 2    Forbidden regions of different stability criteria.

Compared with the Middlebrook criterion [3], the small-gain criterion [21] reduces the forbidden region as the exterior area of the unity circle, and hence it can be regarded as a special case of the Middlebrook criterion without considering any GM. The NSSC [10] is different from other criteria, which reduces the forbidden region as the real axis on the left side of $(-1, j0)$, and it can be regarded as a special case of the GMPM criterion [4] with zero GM and PM. Consequently, the forbidden region becomes a line, and the NSSC provides a necessary and sufficient condition for the stability assessment.

Among those criteria, the Middlebrook criterion, the small-gain criterion, the GMPM criterion, the OPAC, and the NSSC are closely related to the GM and PM of the system. Hence, the stability is usually checked on Bode plots, where the impedance specifications can be implemented. Table II summarizes the various impedance specification methods based on forbidden regions.

In Table II, the Middlebrook criterion and small-gain criterion only specify the magnitudes of the impedances, while the other three criteria specify both magnitudes and phases of the impedances. It is clear that with the aid of Bode plots, the interaction between the numerator and denominator impedances can be investigated, through which the impedances can be better designed and reshaped to enhance the stability of the whole system [22]-[24].

Table II  Impedance specifications based on forbidden regions

| Methods | Forbidden region | Illustration on Bode plots [a] | Features |
|---|---|---|---|
| Middlebrook criterion [3] (sufficient condition) | $\lvert Z_1/Z_2 \rvert > 1/\text{gm}$ <br> ↓ <br> $20\lg\lvert Z_1 \rvert - 20\lg\lvert Z_2 \rvert > -\text{GM}$ | | Impedance specifications on magnitudes, designs for GM |
| Small-gain criterion [21] (sufficient condition) | $\lvert Z_1/Z_2 \rvert > 0$ <br> ↓ <br> $20\lg\lvert Z_1 \rvert - 20\lg\lvert Z_2 \rvert > 0$ | | Impedance specifications on magnitudes, designs for GM |
| GMPM criterion [4] (sufficient condition) | $\lvert Z_1/Z_2 \rvert > 1/\text{gm}$ <br> $\angle Z_1/Z_2 < -180°+\text{PM}$ <br> $\angle Z_1/Z_2 > 180°-\text{PM}$ <br> where $\angle Z_1/Z_2 \in (-180°, 180°]$ <br> ↓ <br> $20\lg\lvert Z_1 \rvert - 20\lg\lvert Z_2 \rvert > -\text{GM}$ <br> $\angle Z_1 - \angle Z_2 < -180°+\text{PM}$ <br> $\angle Z_1 - \angle Z_2 > 180°-\text{PM}$ | | Impedance specifications on magnitudes and phases, designs for GM and PM |
| OPAC [5]-[6] (sufficient condition) | $\text{Real}\{Z_1/Z_2\} < -1/\text{gm}$ <br> ↓ <br> $20\lg\lvert Z_1 \rvert - 20\lg\lvert Z_2 \rvert > -\text{GM}$ <br> $\angle Z_1 - \angle Z_2 < -90°-\phi$ <br> $\angle Z_1 - \angle Z_2 > 90°+\phi$ <br> where $\phi = \arcsin\left\lvert \dfrac{1}{\text{gm}} \cdot \dfrac{Z_2}{Z_1} \right\rvert$ | | Impedance specifications on magnitudes and phases, easy for individual impedance specifications |

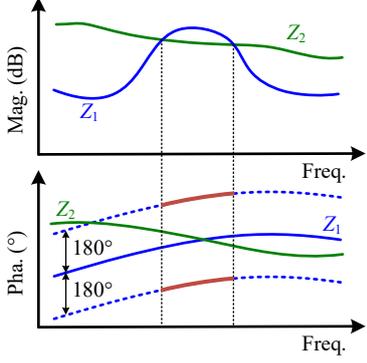

| | | | |
|---|---|---|---|
| NSSC [10] (necessary & sufficient condition) | $\text{Real}\{Z_1/Z_2\} < -1$ $\angle Z_1/Z_2 = 180°$ where $\angle Z_1/Z_2 \in (-180°, 180°]$ ↓ $20\lg|Z_1| - 20\lg|Z_2| > 0$ $\angle Z_1 - \angle Z_2 = 180°$ | | Impedance specifications on magnitudes and phases, accurate stability analysis |

Note: a. red shaded area – forbidden regions.

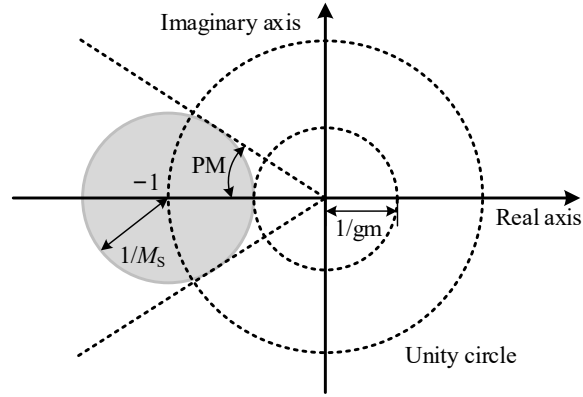

Fig. 3  Forbidden region of the MPC.

Another criterion named as MPC was proposed based on a different forbidden region for the transient performance assessment [9], as shown in Fig. 3. The forbidden region is defined as a disc around $(-1, j0)$ with the radius of $1/M_S$, where $M_S$ is the maximum peak of the sensitivity function $1/(1+ Z_{\text{Type\_Z}}Y_{\text{Type\_Y}})$ [25], as denoted by

$$M_s = \frac{1}{1-1/\text{gm}}. \qquad (5)$$

It is seen that the radius of the forbidden region of MPC is determined by a given GM. Thus, this criterion is merely used for the stability robustness analysis. With this method, the GM and PM can be well designed to enhance the transient performance of the whole system.

*B. Impedance-sum-based method*

Besides Z+Y systems, there are some systems without the Type Y subsystems, such as the HESSs. In a HESS, each subsystem has a stable output impedance, and hence it is a Z+Z system. For Z+Z systems, the RHP zeros may be present in the terminal impedances, thus the RHP poles are likely to be produced in the impedance ratio. To avoid the presence of open-loop RHP poles, the impedance sum criterion was proposed for the stability analysis of Z+Z systems [15].

In a Z+Z system, the impedance sum $Z_1+Z_2$ can serve as the characteristic equation of the interconnected system, according to (1) and (2). The stability can then be determined by the roots of $Z_1+Z_2=0$, namely the zeros of $Z_1+Z_2$. The system will be stable if and only if there are no RHP zeros in $Z_1+Z_2$.

The stability can be checked directly on the zero map of $Z_1+Z_2$. Alternatively, the argument principle can be applied to $Z_1+Z_2$. According to the argument principle [19], in a Z+Z system, there exists

$$\mathcal{N}[Z_1+Z_2] = \mathcal{Z}[Z_1+Z_2] - \mathcal{P}[Z_1+Z_2] = \mathcal{Z}[Z_1+Z_2], \tag{6}$$

where $\mathcal{N}$ denotes the number of Nyquist encirclements around the origin on the complex plane, $\mathcal{Z}$ denotes the number of RHP zeros, and $\mathcal{P}$ denotes the number of RHP poles. Since both $Z_1$ and $Z_2$ have no RHP poles, $\mathcal{P}[Z_1+Z_2]$ equals to zero. The stable condition is that the Nyquist plot of $Z_1+Z_2$ does not encircle the origin.

The impedance sum criterion works as a sufficient and necessary stability condition theoretically, yet it still has some drawbacks:

1) The impedance sum is analyzed as a whole, the interactions between $Z_1$ and $Z_2$ cannot be investigated, which provides little guidance for the reshaping the impedances of subsystems.

2) For a given transfer function, its frequency response will approach to zero at $\omega=0$ or $\omega=\infty$, as long as the numerator has a different degree from the denominator. Therefore, the frequency response at $\omega=0$ and $\omega=\infty$ are critical for determining the number of Nyquist encirclements around the origin. However, for a specific system with power electronic converters, even though the frequency response of $Z_1+Z_2$ can be obtained, the frequency response at $\omega=0$ and $\omega=\infty$ may not be available. If the impedance is obtained analytically, it is usually valid below the half switching frequency of converters, and then the frequency response at $\omega=\infty$ has no meaning. On the other hand, if the impedance is measured by frequency scanning, it is also impossible to obtain the frequency response at $\omega=0$ and $\omega=\infty$, and the obtained Nyquist plot is not a closure. Therefore, it is not convincing to use the argument principle for the stability assessment in practice.

*C. NSC for multi-loop systems*

Beyond the impedance sum criterion, another solution to prevent the presence of open-loop RHP poles is to apply the NSC for multi-loop systems, which was formulated by Bode as [26], [27]:

"When a linear system is stable with certain loops disconnected, it is stable with these loops closed if and only if the total numbers of clockwise and counterclockwise encirclements of the point (−1,0) are equal to each other in a series of Nyquist diagrams drawn for each loop and obtained by beginning with all loops open and closing the loops successively in any order, leading to the system normal configuration."

This method can be used for a paralleled converter system, since it can be seen as a multi-loop system [17], as shown in Fig. 4(a). The source and the impedance $Z_0$ are regarded as the original system. Each connection of a paralleled converter can be regarded as an additional loop, whose model can be represented by Fig. 4(b). With all loops open, the system is always stable, such that the impacts caused by open-loop RHP poles can be avoided. Then the stability can be assessed by a series of Nyquist plots of $Z_{k,c}Y_{k+1}$. If the total number of encirclements of Nyquist plots of $Z_{k,c}Y_{k+1}$ (for all $k$) around the critical point equals to zero, the system is stable.

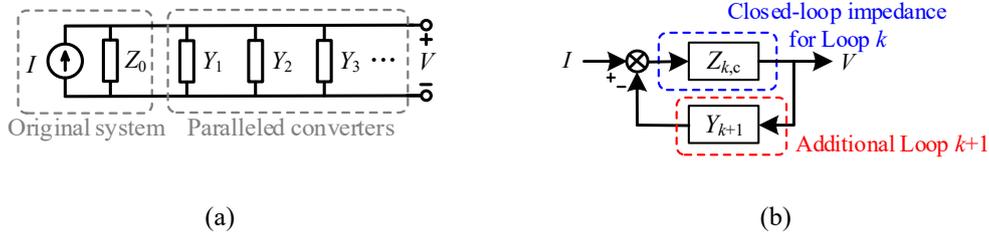

Fig. 4  A paralleled converter system. (a) Equivalent circuit; (b) Multi-loop model representation.

Although the NSC for multi-loop systems works well theoretically, the calculation is burdensome since it requires a series of Nyquist plots for the stability analysis. Moreover, how to select the order of closing the loops and whether it has physical insights or not remain elusive.

III. GENERAL RULE FOR IMPEDANCE-BASED STABILITY ANALYSIS

In this section, a general rule of using the impedance-ratio for the stability analysis is proposed, which is derived from the NSC, leading to a sufficient and necessary stability condition, and meanwhile, enabling to specify impedance profiles for a design-oriented analysis. The stability analysis consists of four stages, the formulation of the impedance ratio, the identification of the open-loop RHP poles, the identification of the encirclements, and the stability analysis.

*A. Formulation of the impedance ratio*

It has been discussed in the previous section that how to formulate the impedance ratio plays a significant role in the stability analysis. The numerator and denominator impedances can be chosen according to the types of subsystems (Type Z or Y) to avoid the presence of open-loop RHP poles. However, this method may not apply to Z+Z systems, since the open-loop RHP poles can be produced by the ratio calculation if the denominator impedance has RHP zeros. Hence, the prevention of open-loop RHP poles by defining the impedance ratio is not an effective way.

It is known that the NSC assumes that the open-loop transfer function is a proper function [18], [19], which implies that the numerator has lower degree than the denominator, or the frequency response of the transfer function approaches to the origin at the infinity frequency. However, for an artificially defined impedance ratio, this condition may not hold. Then the NSC is not straightforward for the stability analysis, due to the non-zero value of the frequency response at the infinite frequency, and the inverse NSC is a better choice [20]. Hence, in this paper, it is suggested to define the impedance ratio as a proper transfer function, named as the proper impedance ratio, which satisfies

$$\lim_{\omega \to \infty} \frac{Z_1(j\omega)}{Z_2(j\omega)} = 0 . \qquad (7)$$

Eq. (7) also implies that the impedance ratio is formulated as the "minor" loop gain, since $|Z_1| < |Z_2|$ at the infinite frequency. For the cases of input filter designs for dc-dc converters, source-load impedance ratio for DPSs, load-source impedance ratio for CSCSs, and the defined impedance ratio in (3) for Z+Y systems, all these impedance ratios are formulated as proper impedance ratios. Such definition is applicable for all kinds of systems, also including the Z+Z systems.

It should be mentioned that in a practical system, the frequency response at the infinite frequency may not be available. The formulation of the impedance ratio can be determined by the given high-frequency responses and their derivatives of the two impedances, since the tendency of frequency response of each impedance approaching to the infinite frequency can be estimated. At higher frequencies, if the two impedances have the different magnitude slopes, the numerator impedance should be chosen as the one with the smaller magnitude slope. If the two impedances have the same magnitude slope at higher frequencies, the numerator impedance should be chosen as the one with the smaller magnitude.

## B. Identification of the open-loop RHP poles

As the impedance ratio is intentionally defined as a proper function, the RHP poles may be produced in the impedance-ratio, if there are non-minimum-phase subsystems [19]. The number of open-loop RHP poles has to be known prior to the stability analysis. One method for identifying RHP poles or zeros is based on the parametric identification of the impedance frequency response [28], and then the fitted impedance transfer function is checked with the pole-zero map. Another more intuitive method is to identify the RHP poles and zeros directly from the impedance frequency response, which is simpler and is adopted in this paper. Table III shows the identification rule of RHP poles and zeros with the aid of Bode plots [19], which presents opposite characteristics of left-half-plane poles and zeros for the minimum-phase systems.

Given the frequency responses of $Z_1$ and $Z_2$, the number of RHP poles in the numerator impedance $Z_1$ and the number of RHP zeros in the denominator impedance $Z_2$ can be identified, and the number of open-loop RHP poles is then calculated as

$$\mathcal{P}[Z_1/Z_2] = \mathcal{P}[Z_1] + \mathcal{Z}[Z_2]. \tag{8}$$

Table III  Identification rule of RHP poles and zeros

| Types | Identification rule (at break points) | Illustration on Bode plots |
|---|---|---|
| RHP real poles | Magnitude: slope change of −20dB/dec<br>Phase: step change of +90° | (Bode plot showing magnitude with 0dB/dec transitioning to −20dB/dec, and phase with +90° step at $f_b$) |
| RHP real zeros | Magnitude: slope change of +20dB/dec<br>Phase: step change of −90° | 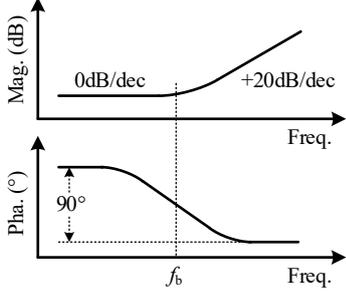 |
| RHP conjugate poles | Magnitude: slope change of −40dB/dec<br>(with a resonant peak if $\zeta < 0.7$)<br>Phase: step change of +180° | 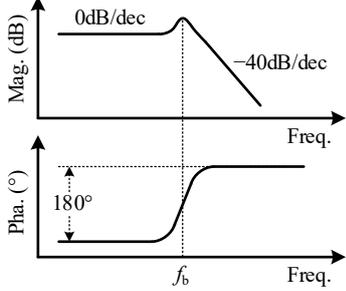 |
| RHP conjugate zeros | Magnitude: slope change of +40dB/dec<br>(with an anti-resonant peak if $\zeta < 0.7$)<br>Phase: step change of −180° | 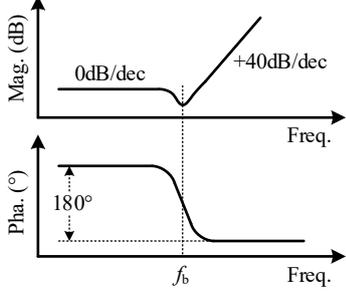 |

Notes: $\zeta$ – damping ratio for the second-order term, $f_b$ – break point.

*C. Identification of the encirclements*

According to the NSC, the system is stable if and only if

$$\mathcal{N}\,[Z_1/Z_2] = -\mathcal{P}\,[Z_1/Z_2]. \tag{9}$$

The $\mathcal{P}\,[Z_1/Z_2]$ can be obtained by (8) on the Bode plots of $Z_1$ and $Z_2$. It will be easier for analysis if the $\mathcal{N}\,[Z_1/Z_2]$ can be directly determined on the same plot. A graphical method using Bode plots to identify the number of encirclements is introduced below.

Fig. 5 maps the encirclements around the critical point from the Nyquist plot to the Bode plot. On the Nyquist plot of $Z_1/Z_2$, the number of encirclements around the critical point $(−1, j0)$ can be determined by the number of crossings over $(-\infty, -1)$ on the real axis, as denoted by the $R$ in Fig. 5 (a). There may be two types of crossings, i.e. one type is the clockwise crossing (CC), and the other is the anti-clockwise crossing (ACC). Thus the total number of crossings can be formulated as

$$\mathcal{N}[Z_1/Z_2] = \mathcal{N}_{CC}[Z_1/Z_2] - \mathcal{N}_{ACC}[Z_1/Z_2], \qquad (10)$$

where both $\mathcal{N}_{CC}[Z_1/Z_2]$ and $\mathcal{N}_{ACC}[Z_1/Z_2]$ are non-negative integers. They can be obtained by

$$\mathcal{N}_{CC}[Z_1/Z_2] = \left\| \left\{ \omega \in \mathbb{R} \mid \left|\frac{Z_1}{Z_2}(j\omega)\right| > 1 \text{ and } \angle \frac{Z_1}{Z_2}(j\omega) = \pm 180° \text{ and } \frac{d}{d\omega}\frac{Z_1}{Z_2}(j\omega) < 0 \right\} \right\|, \quad (11)$$

$$\mathcal{N}_{ACC}[Z_1/Z_2] = \left\| \left\{ \omega \in \mathbb{R} \mid \left|\frac{Z_1}{Z_2}(j\omega)\right| > 1 \text{ and } \angle \frac{Z_1}{Z_2}(j\omega) = \pm 180° \text{ and } \frac{d}{d\omega}\frac{Z_1}{Z_2}(j\omega) > 0 \right\} \right\|. \quad (12)$$

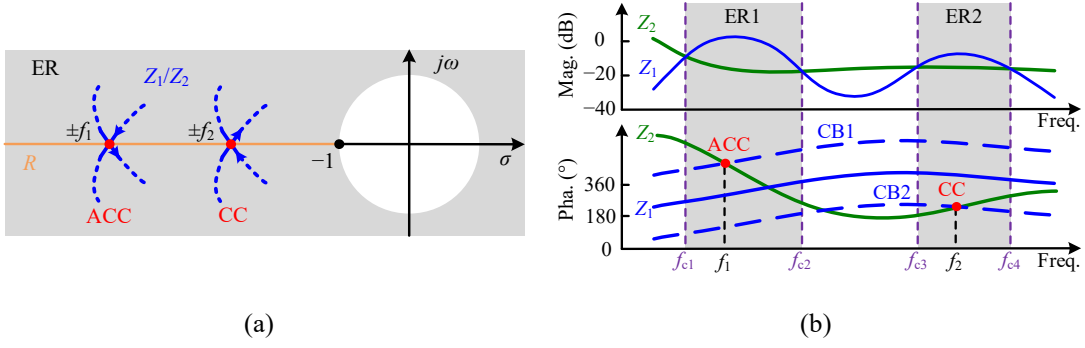

Fig. 5  Illustration of the number of encirclements around the critical point. (a) Nyquist plot; (b) Bode plot.

It is noticed from (11) and (12) that the direction of the Nyquist trajectory is related to the derivative of the frequency response of $Z_1/Z_2$. For the conventional stability analysis methods with Bode plots, only the values of the frequency response (magnitude and phase) are considered, yet the derivatives of the frequency response are overlooked [29]. It is assumed that there are one ACC at $f_1$ and one CC at $f_2$. From the Nyquist plot of Fig. 5(a), the crossing type at $-f_1$ should be the same as that at $f_1$, and similarly for $\pm f_2$. Formulating (11) and (12) on the Bode plot, Fig. 5(b) is obtained according to the rule proposed in Table IV, which consists of four steps. With such method, it is easy to obtain the number of encirclements directly from the impedance Bode plots of $Z_1$ and $Z_2$ without doing the ratio calculation and drawing the Nyquist plot.

For Step 1, the exterior regions (ERs) outside the unity circle are found as all the frequency intervals where the numerator impedance ($Z_1$) is larger than the denominator impedance ($Z_2$), as denoted by the shaded areas in Fig. 5.

For Step 2, two auxiliary boundaries by shifting one impedance-phase curve (i.e. $\angle Z_1$) of $\pm 180°$ are drawn to help identify the crossings, which are defined as two crossing boundaries (CBs). If the other impedance-phase curve (i.e. $\angle Z_2$) intersects with the two CBs within the ERs, there will be crossings over $R$ on the Nyquist plot, as denoted by the red dots in Fig. 5.

For Step 3, the crossing types can be determined by the derivatives of the impedance frequency responses. At each crossing frequency, if the phase derivative difference between the numerator impedance ($Z_1$) and the denominator impedance ($Z_2$) is negative, the crossing type will be a CC, and vice versa for an ACC.

Table IV  Identification rule of the number of encirclements on Bode plots

| | Steps | Mathematical representation |
|---|---|---|
| Step 1 | Find all the ERs, as denoted by the shaded areas. | $\|Z_1(j\omega)\| > \|Z_2(j\omega)\|$ |
| Step 2 | Check whether there are crossings within the ERs with the aid of the CBs, as denoted by the red dots. | $\angle Z_1(j\omega) - \angle Z_2(j\omega) = \pm 180°$ |
| Step 3 | Determine the crossing types if the crossings exist. | $\frac{d}{d\omega}Z_1(j\omega) < \frac{d}{d\omega}Z_2(j\omega)$ for a CC $\frac{d}{d\omega}Z_1(j\omega) > \frac{d}{d\omega}Z_2(j\omega)$ for an ACC |
| Step 4 | Calculate the number of crossings. | $N_{CC}[Z_1/Z_2] = 2n_{cc} + n_{cc0}$ [a] $N_{ACC}[Z_1/Z_2] = 2n_{acc} + n_{acc0}$ [b] $N[Z_1/Z_2] = N_{CC}[Z_1/Z_2] - N_{ACC}[Z_1/Z_2]$ |

Notes: a. $n_{cc}$ – number of CCs for $\omega \neq 0$, $n_{cc0}$ – number of CCs at $\omega = 0$.
      b. $n_{acc}$ – number of ACCs for $\omega \neq 0$, $n_{acc0}$ – number of ACCs at $\omega = 0$.

For Step 4, the total number of crossings can be calculated by the given formulas in Table IV. Since the crossing types are the same for a positive frequency and its inverse, the number of crossings should be doubled for $\omega\neq0$. For $\omega=0$, there is only one crossing, hence the number of crossings should not be doubled. The total number of encirclements can be calculated by (10).

*D. Stability analysis*

Given the number of open-loop RHP poles and the number of encirclements, the NSC can be applied with (8)-(10). The system will be stable if and only if Eq. (13) holds.

$$\mathcal{N}_{\mathrm{CC}}\,[Z_1/Z_2] - \mathcal{N}_{\mathrm{ACC}}\,[Z_1/Z_2] = -\,\mathcal{P}\,[Z_1] - \mathcal{Z}\,[Z_2]. \tag{13}$$

The proposed rule implemented on Bode plots is equivalent to the NSC. Among the four steps in Table IV, Steps 1-2 suggest the principle of the conventional stability analysis methods, which only considers the magnitude and phase values of the frequency response. While Steps 3-4 indicate the new procedure of the proposed rule, wherein the derivatives of the frequency response is considered in Step 3. Thus, the direction of the Nyquist trajectory can be considered in the analysis, and the crossing number and types can be easily identified on Bode plots by Step 4.

*E. Advantages of the proposed stability analysis method*

Table V gives a comprehensive comparison of the proposed stability analysis method and the conventional ones. In contrast to the conventional methods, the main advantages possessed by the proposed method can be summarized as follows:

1) The proposed stability analysis method based on the proper impedance ratio (open-loop gain) is more general. It is applicable for both Z+Y systems and Z+Z systems, since the open-loop RHP poles are considered in the analysis.

Table V  Comparison of different impedance-based stability analysis methods

| Stability analysis methods | Forbidden region | NSC | Impedance sum | Multi-loop NSC | Proposed |
|---|---|---|---|---|---|
| Analysis basis | Open-loop gain | Open-loop gain | Characteristic equation | Multiple open-loop gains | Open-loop gain |
| Application | Z+Y systems | All systems | Z+Z systems | All systems | All systems |
| Tools required | Bode plot | Pole-zero map Nyquist plot | Nyquist plot | Multiple Nyquist plots | Bode plot |
| Calculation | Less | Impedance ratio | Impedance sum | Multiple open-loop gains | Less |
| Impedance specification | Yes | No | No | No | Yes |

Notes: "Less" in the "Calculation" row means neither the impedance ratio nor the impedance sum is required for calculation.

2) Given the impedance frequency responses, it is available to identify the number of RHP poles or zeros from the Bode plot without drawing the pole-zero map. There is also no need to calculate the impedance ratio throughout the analysis. The stability can be readily predicted by the four-stage rule based on the Bode plot, such that the analysis is easier and more intuitive.

3) The stability analysis on Bode plots is more design-oriented than that on Nyquist plots. Even for cases with open-loop RHP poles, in order to guarantee the stability, the impedance can be specified to cross over the CBs in the ERs the same times as the number of open-loop RHP poles. This advantage will be further illustrated in the next section.

IV. Experimental Case Studies

The stability analyses and experiments on a paralleled inverter system are carried out in this section to validate the proposed method. The system configuration is displayed in Fig. 6.

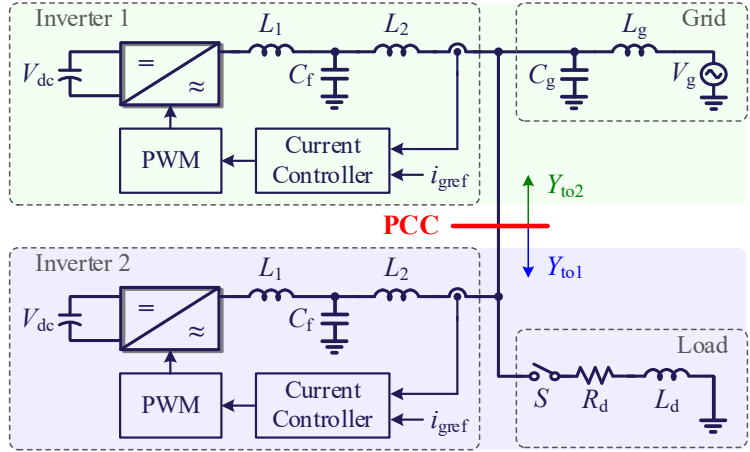

Fig. 6   The paralleled inverter system.

Table VI   System parameters

| Parameters | Values | Parameters | Values |
| --- | --- | --- | --- |
| $V_{dc}$ | 730V | $L_1$ | 1.8mH |
| $V_{grms}$ (line to line) | 400V | $L_2$ | 0.9mH |
| $L_g$ | 1mH | $C_f$ | 10μF |
| $C_g$ | 2μF | $K_p$ | 8 |
| $R_d$ | 10Ω | $K_r$ | 500 |
| $L_d$ | 1mH | $f_s$ | 10kHz |

Notes: a,b. $K_p$ and $K_r$ – parameters of the current-loop PR controller.

c. $f_s$ – sampling frequency which is also equal to the switching frequency.

Two inverters are connected to the grid, and the grid-side current control is implemented with the proportional-resonant (PR) controller. The load is selected by the switch $S$, and thus two operating scenarios are considered. For Scenario I, $S$ is switched off. The system can be seen as a paralleled grid-tied inverter system, which is usually found in the wind or solar photovoltaic systems. For Scenario II, $S$ is switched on. The system can be regarded as a distributed power system. All the parameters are selected as shown in Table VI. Two inverters have identical parameters.

*A. System modeling*

Since the grid-tied inverter in Fig. 6 operates with a stable output admittance, and it is easier to aggregate different subsystems by adding the admittances up, the admittance ratio is adopted for the analysis. The inverter admittance can be modeled based on the small-signal model of the inverter in Fig. 7 [30].

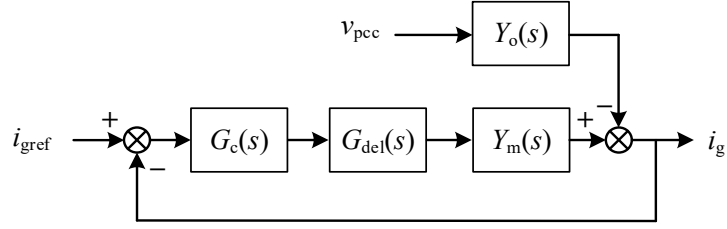

Fig. 7    Small-signal model of the inverter.

In Fig. 7, the admittances $Y_o$ and $Y_m$ can be obtained by [30]

$$Y_o(s) = \frac{Z_{L1} + Z_{Cf}}{Z_{Cf}Z_{L1} + Z_{L1}Z_{L2} + Z_{Cf}Z_{L2}}, \tag{13}$$

$$Y_m(s) = \frac{Z_{Cf}}{Z_{Cf}Z_{L1} + Z_{L1}Z_{L2} + Z_{Cf}Z_{L2}}. \tag{14}$$

The PR controller is used for the current loop control [31], whose transfer function is

$$G_c(s) = K_p + \frac{2K_r \omega_c s}{s^2 + 2\omega_c s + \omega_1^2}, \tag{15}$$

where $K_p$ and $K_r$ are the P and R parameters, $\omega_1$ is the fundamental frequency, and $\omega_c$ is selected as 3.14 rad/s to slightly widen the bandwidth of the PR controller [31].

$G_{del}$ denotes the transfer function of the delay, which can be expressed by the Pade approximation [17]. In this paper, a third-order Pade approximation is used, which leads to

$$G_{del}(s) = Pade(e^{-1.5T_s s}, 3) = \frac{1 + \frac{1}{2}(-1.5T_s s) + \frac{1}{8}(-1.5T_s s)^2 + \frac{1}{48}(-1.5T_s s)^3}{1 - \frac{1}{2}(-1.5T_s s) + \frac{1}{8}(-1.5T_s s)^2 - \frac{1}{48}(-1.5T_s s)^3}. \tag{16}$$

In the inverter model, the impacts of the phase-locked loop is neglected by selecting a sufficiently low control bandwidth for synchronizing with the grid. Thus the model of Fig. 6 can be represented in the single-input single-output form [32]. The derived inverter output admittance is

$$Y_{io}(s) = \frac{Y_o(s)}{1 + G_c(s)G_{del}(s)Y_m(s)}. \tag{17}$$

The grid admittance can be easily deduced as

$$Y_g(s) = sC_g + \frac{1}{sL_g}, \tag{18}$$

and the load impedance is

$$Y_d(s) = \frac{1}{R_d + sL_d}. \tag{19}$$

For the stability analysis, the admittance ratio should be defined first, which requires the whole system to be divided as two subsystems. In the analysis of this paper, the Inverter 1 and the grid are regarded as one subsystem, which bring RHP zeros in the aggregated admittance, i.e. $Y_{to2}$. This aggregation is intentionally for validation. The rest of the system is regarded as the other subsystem, whose admittance is denoted as $Y_{to1}$. Therefore, the PCC is selected as shown in Fig. 6.

*B. Stability analysis of Scenario I*

For Scenario I, the load is switched off. $Y_{to1}$ is the admittance of the Inverter 2, and $Y_{to2}$ is the aggregated admittance of the Inverter 1 and the grid. The admittance Bode plots of $Y_{to1}$ and $Y_{to2}$ are shown in Fig. 8. It can be seen that $|Y_{to2}| \gg |Y_{to1}|$ at high frequencies, thus it is better to use the admittance ratio $Y_{to1}/Y_{to2}$ for stability analysis according to (7).

However, from the pole-zero map of $Y_{to1}$ and $Y_{to2}$ in Fig. 9, two pairs of conjugate RHP zeros can be observed in $Y_{to2}$. Thus, there are four RHP poles produced in the admittance ratio $Y_{to1}/Y_{to2}$. According to the conventional stability analysis methods, the Nyquist plot of $Y_{to1}/Y_{to2}$ has to be checked due to the presence of open-loop RHP poles, as shown in Fig. 10. It can be found that the Nyquist trajectory does not encircle the critical point, which implies that the system is unstable. Although the Nyquist plot can predict the stability, the information of open-loop RHP poles cannot be directly obtained from Fig. 10, and the analysis result provides little insight into impedance interactions and little guidance for shaping the impedance.

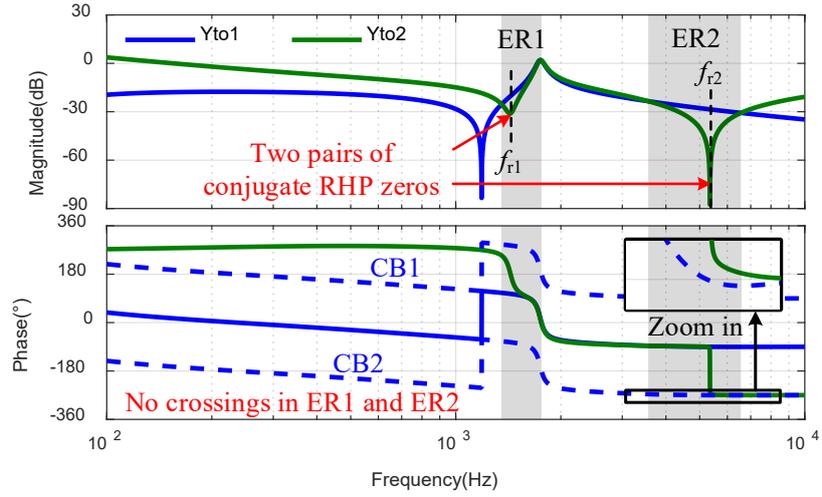

Fig. 8    Admittance Bode plot for Scenario I.

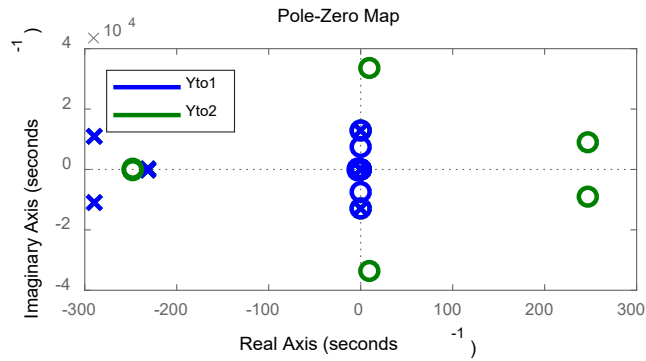

Fig. 9    Admittance pole-zero map for Scenario I.

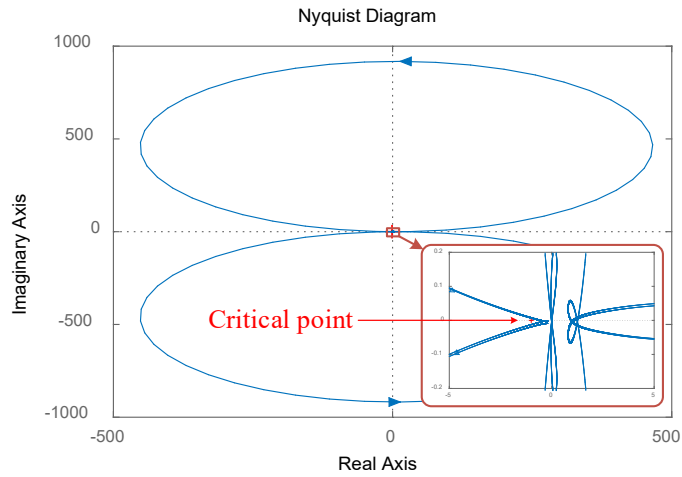

Fig. 10    Admittance-ratio Nyquist plot for Scenario I.

With the proposed method, the system stability can be predicted directly from the admittance Bode plot, without calculating the admittance ratio. The RHP poles in $Y_{to1}$ and the RHP zeros in $Y_{to2}$ should be checked first. It can be seen from Fig. 8 that there are two anti-resonances at $f_{r1}$ and $f_{r2}$ in the magnitude plot of $Y_{to2}$, where the corresponding phase changes are $-180°$. This fact implies two pairs of conjugate RHP zeros at $f_{r1}$ and $f_{r2}$, and hence there are four RHP poles present in the admittance ratio $Y_{to1}/Y_{to2}$, which agrees with the result from Fig. 9.

In Fig. 8, there are two ERs where $|Y_{to1}| > |Y_{to2}|$, as denoted in the shaded areas. The two CBs are drawn by shifting the phase curve $\angle Y_{to1}$ of $\pm 180°$. Within the ERs, there is no crossing between the CBs and the phase curve $\angle Y_{to2}$, which indicates that there is no encirclement around the critical point for the Nyquist trajectory of $Y_{to1}/Y_{to2}$. The analysis result is the same as that obtained from Fig. 10. Hence, the total number of anticlockwise encirclements is not equal to the number of open-loop RHP poles, and the system is unstable.

The experimental waveforms for the Scenario I are presented in Fig. 11. The waveform in CH1 is the line-to-line voltage at the PCC, and the waveforms in CH2-CH4 are the output currents for the Inverter 1. It can be seen that the system is unstable, which is predicted by the stability analysis.

The case study of Scenario I shows that the proposed method using Bode plots works well in the stability analysis even for cases with the open-loop RHP poles. It is more intuitive for identifying the number of RHP pole/zero and requires less calculation than drawing the Nyquist plot.

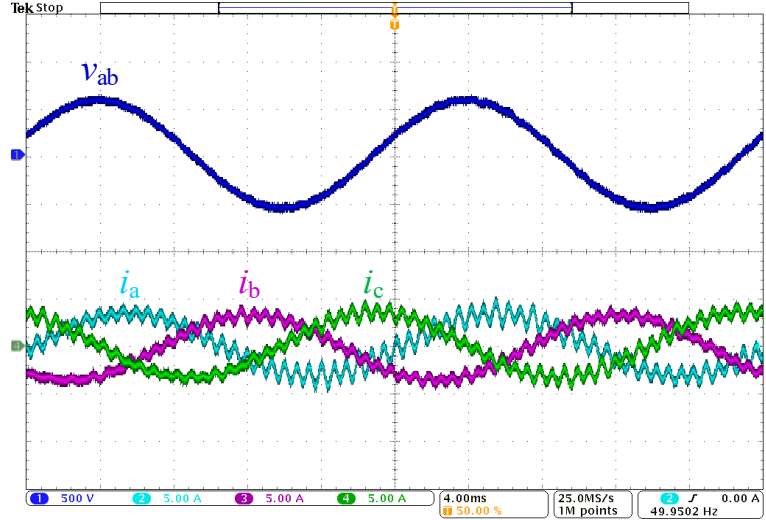

Fig. 11  Experimental waveforms for Scenario I.

*C. Stability analysis of Scenario II*

It is seen from Fig. 8 that the system is unstable for Scenario I, since there is no crossing within the ERs. One way to stabilize the system is to specify the admittance $Y_{to1}$ by letting the CBs of $\angle Y_{to1}$ crosses over $\angle Y_{to2}$ two times anticlockwise within the ERs (corresponding to four ACCs including the negative frequency range).

For Scenario II, $Y_{to2}$ does not change compared with Scenario I. When the load is connected to the system, $Y_{to1}$ becomes the aggregated admittance of the Inverter 2 and the load, so $Y_{to1}$ is reshaped as shown in Fig. 12. It is clear that $|Y_{to2}|$ is still much larger than $|Y_{to1}|$ at high frequencies, and hence the admittance ratio $Y_{to1}/Y_{to2}$ is adopted. The two CBs of the reshaped admittance $Y_{to1}$ are drawn in the same way, which are now crossing the phase curve of $Y_{to2}$ two times within the ER, as denoted by the red dots. At each crossing point, the derivative of $\angle Y_{to1}$ is larger than that of $\angle Y_{to2}$, indicating an ACC. Consequently, there are two ACCs within the ER, and the total number of ACCs is four including the negative frequency range, which equals to the number of open-loop RHP poles. Hence, the system becomes stable with the load connected.

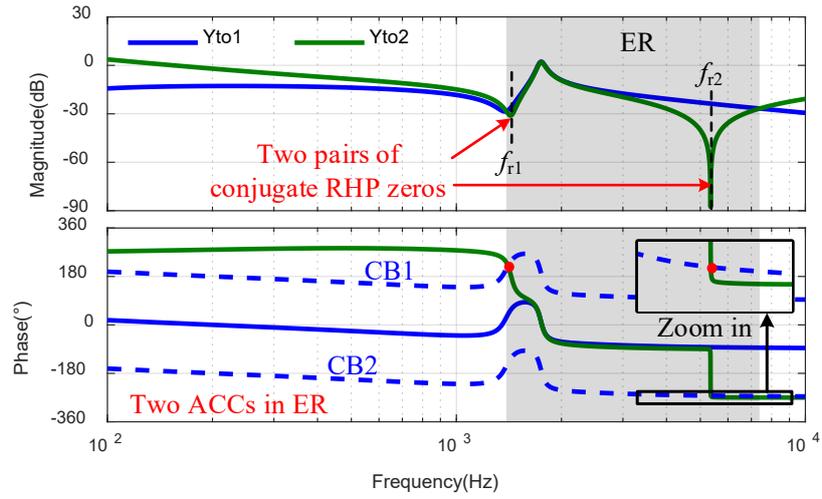

Fig. 12 Admittance Bode plot for Scenario II.

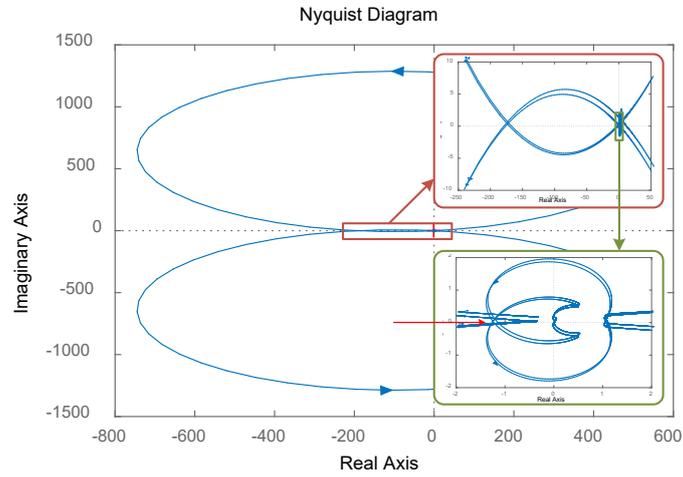

Fig. 13 Admittance-ratio Nyquist plot for Scenario II.

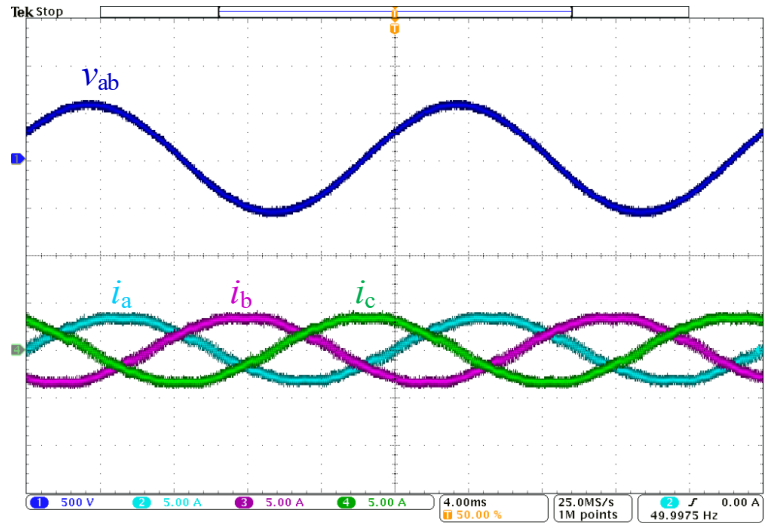

Fig. 14 Experimental waveforms for Scenario II.

Fig. 13 shows the stability analysis result by drawing the Nyquist plot of the admittance ratio, where four anticlockwise encirclements around the critical point are observed, which also verifies the correctness of the proposed method. However, from both Fig. 10 and Fig. 13, it is hard to obtain further insights into the interactions of subsystems. Fig. 14 shows the stable experimental waveforms for Scenario II, which is in accordant with the stability analysis result.

The study of Scenario II shows that the stability analysis on Bode plots is more design-oriented, which can provide some guidance for system stabilizations and impedance specifications, even when the RHP poles are present in the impedance ratio.

## V. Conclusion

This paper has put forward a general impedance-based stability analysis method implemented on Bode plots. The derivatives of the frequency response versus frequency, which was ignored in the conventional analysis, is considered in the proposed method. Two case studies in experiments have demonstrated the effectiveness and superior features of the proposed method. One advantage is the easier implementation, since it is not necessary to draw the pole-zero map and calculate the impedance ratio. The other advantage is that the analysis on Bode plots is more design-oriented, and the impedance can be specified to enhance the system stability, even for systems with open-loop RHP poles.